\documentclass{andp2012}

\usepackage{graphicx}
\usepackage[numbers,sort&compress]{natbib}
\usepackage{hyperref}
\usepackage{amsmath}
\usepackage{mathrsfs}
\usepackage{amsfonts}
\usepackage{color}
\usepackage{bm} 
\usepackage{subfigure}
\usepackage{epsfig} 
\usepackage{epstopdf}
\usepackage{float} 
\usepackage{tikz}
\usetikzlibrary{shapes.arrows}
\usepackage[percent]{overpic}
\usepackage{comment}
\usepackage{pstricks}
\usepackage{filecontents}
\usepackage{enumitem}
\usepackage{booktabs}
\usepackage{atbegshi}
\AtBeginDocument{\AtBeginShipoutNext{\AtBeginShipoutDiscard}}

\renewcommand{\thesection}{\arabic{section}}

\shortabstract
\begin{document}

\title{ Deterministic secure quantum communication with and without entanglement}

\author{Tarek A. Elsayed}
\email{tarek.elsayed@aucegypt.edu}
\address{Department of Physics, School of Science and Engineering, The American University in Cairo, AUC Avenue, P.O. Box 74, New Cairo, 11835, Egypt}
\address{Department of Physics and Astronomy, Hunter College of the City University of New York,New York 10065, USA}



\clearpage
\setcounter{page}{1}
\begin{abstract}

We present a protocol for sending a message over a quantum channel with different layers of security that will prevent an eavesdropper from deciphering the message without being detected.  The protocol has two versions where the bits are encoded in either pairs of entangled photons or separate photons. Unlike many other protocols, it requires a one-way, rather than a two-way, quantum channel and does not require a quantum memor. A quantum key is used to encrypt the message and both the key and the message are sent over the quantum channle with the same quantum encoding technique. The key is sent only if no eavesdropper is detected.


\end{abstract}
\maketitle

\section{Introduction}

Quantum cryptography has become one of the most fruitful and versatile commercial applications of quantum information. While classical encryption can in principle be compromised with a powerful enough computer, quantum encryption provides a platform where any eavesdropping attempt can be detected with a very high probability. There are several major schemes where quantum encryption is employed such as: (i) Quantum key distribution (QKD) where a random key is generated and securely shared between two parties and used later in classical encryption.  (ii) Quantum secure direct communication (QSDC) where a certain message is securely and directly transferred between two parties using a quantum algorithm without the need for sharing a secure key or sending data over the classical channel except for detecting an eavesdropper. (iii) Deterministic secure quantum communication  (DSQC) where the message is also sent deterministically over a quantum channel with the help of sending data over the classical channel \cite{long2007}. While the experimental realization of QKD has been achieved at least as early as 1992 \cite{bennett1992}, the proof-of-concept experiments of quantum direct communication has been achieved only recently \cite{hu2016, sun2018,zhu2017,qi2019}.

There is a wide variety of proposals for each of these schemes that differ in terms of the states of the photons used (entangled photons or single photons) and the type of the quantum channel (one-way or two-way channel). While the oldest QKD protocol (BB84) introduced by Bennett and  Brassard in 1984 \cite{bb84} uses single photons, there are other protocols that use entangled pairs of photons \cite{ekert91,bennett1992}. Similarly, there are numerous QSDC schemes that use entangled photons, usually in one of Bell states (\cite{zhu2006,bostrom2002,deng2003, cai2004,gao2019,li2020,zhou2020,zhou2020-2,xie2020}) and others which use single photons \cite{deng2004,hu2016}.
DSQC schemes also use either entangled photons \cite{shimizu1999,li2006,joy2017} or single photons \cite{beige2001,li2006,huang2012,jiang2017}. While there exists direct quantum communication protocols which require a one-way quantum channel such as \cite{deng2003},  many QSDC/DSQC  protocols require two-way quantum channels where photons are sent back and forth between Alice and Bob (the two famous parties who know the laws of quantum mechanics very well and use them in order to secure their communication). This later type requires storing the qubits for a long time using a quantum memory which may be difficult to achieve due to their short coherence time and requires also the precise control of the timing of their manipulation. Overcoming these difficulties by implementing an atomic quantum memory \cite{zhang2017} or devising a new protocol that does not require a quantum memory \cite{sun2018} was only achieved very recently.

 QKD can be implemented by sending single photons using only a single degree of freedom, i.e., using a two-dimensional Hilbert space, as in BB84 protocol. For sending data in a deterministic manner using a one-way quantum channel, we need at least a four-dimensional Hilbert space \cite{beige2001,beige2002}. For example, in the protocol proposed by A. Beige et. al. \cite{beige2002} both the spatial and polarization degrees of freedom of single photons are used.
In DSQC/QSDC, we not only aim to detect an eavesdropper (let us call him Evan) with a very high probability, but also to prevent Evan from discerning a good part of the message before being detected \cite{deng2004}. One of the main ideas in this paper is that fulfilling the first aim actually facilitates the fulfillment of the second one, by sending an encrypted message while sending the key to decrypt this message only after the safety of the communication channel is verified. This can be done in several ways. For example, we can preprocess the message before sending it with a DSQC protocol using a symmetric cryptographic algorithm and send the crypto-key (using a similar DSQC protocol) only if the channel is safe. In this way,  we can ensure that even if Evan discerned any part of the sent packet, he will not be able to decipher the message since the key will not be available to him. Another method to fulfill this aim is simply to shuffle the bits constituting the message in a random order and only send the  information used to restore the order of each bit after ensuring the privacy of the channel.

In this paper, we present a scheme where both the  key and the encrypted message are sent over a quantum channel using pairs of single photons or entangled photons.   The two protocols require a one-way quantum channel in addition to the  classical channel and use a similar pre- and postprocessing of the transmitted bits (Figure 1) but differ in the quantum encryption part (Figure 2). In section 2, we present the first protocol using unentangled photons and describe the classical preprocessing common in the two protocols. In section 3, we present the second protocol using entangled photons. The two protocols are described using generic quantum circuits. Finally, in section 4, we analyze the robustness of these protocols against famous eavesdropping schemes.

\section{DSQC protocol without entanglement}

In this protocol, Alice encodes each bit by two photons (we will refer to photons as qubits henceforth) encoded in two different bases assigned to the two qubits randomly. For general qubits, a Hadamard gate (H) can be inserted to one of the two qubits selected randomly after being encoded in the computational basis with the state of the classical bit. Therefore,  `1' is encoded by either of the two-qubit states $|+1\rangle$ or $|1+\rangle$ and similarly `0' is encoded by either $|-0\rangle$ or $|0-\rangle$ randomly selected,  where $|+\rangle=\frac{1}{\sqrt{2}}(|0\rangle+|1\rangle)$ and $|-\rangle=\frac{1}{\sqrt{2}}(|0\rangle-|1\rangle)$. In the case for photons, the two bases can be the rectilinear and diagonal polarization.  Bob, on the other hand, measures the two qubits always in the same basis which can be either one of them randomly for each pair (see Figure 2-a, b). By doing so, and assuming a noiseless channel,  he ensures that at least one of the two qubits will be measured in the correct basis. The measurement outcome of the other qubit will be completely random. In cases where his measurements of the two qubits agree, he knows for sure which bit was encoded by Alice without the need for classical communication. For the other cases, Bob will send to Alice over the classical channel the locations of the pairs where his measurement outcomes are different. Alice, in turn, will send him over the classical channel her choices for these cases.  Bob will then find out which one of the two qubits was measured without passing through a Hadamard gate (H) in either side or with passing through H in both sides. In both cases, the measurement outcome of this bit is the true classical bit encoded by Alice since $\text{HH}=\mathbb{1}$

So far, we have introduced only the  quantum encryption part of our DSQC protocol.  In order to detect eavesdropping and ensure that Evan cannot decode any part of the message before he is detected, more layers of complexity should be added at each level (see Figure 1). For example, Alice can insert a random subset of bits (redundancy check bits) into the main message at random locations and communicate with Bob in public at the end of transmission her choices for these bits together with their locations. An eavesdropper intervening in the middle by doing any kind of measurements will spoil the encoding of the redundant qubit pairs. Moreover, in order to prevent Evan from detecting any sequence of bits before being detected, the packet is encrypted with some sort of symmetric-key encryption algorithms before being sent to Bob. The key is generated at Alice's side and sent to Bob in the same manner at the end of the encrypted packet transmission only if no eavesdropping is detected. Consequently, even if Evan could intercept the entire encrypted message by posing as Bob,  he would not be able to get any useful information from it without the key used by Alice to encrypt the message. In other words, in order for Evan to get any part of the packet he needs to know both the exact key and the exact encrypted message without being detected which is very improbable to happen. 
 
 Let us now outline the complete algorithm in detail. 
 
 \begin{enumerate}
 
 \item Alice divides the full message into small packets $M$, and computes a hash value $S$ for each packet, such as a cyclic redundancy check (CRC) \cite{peterson1961} or a checksum to detect errors in the transmission. Let us denote each of the new packets resulting after appending $S$ to $M$ as $C$. 
\item Alice generates a random key $K$ and use it to encrypt $C$ by a symmetric key algorithm \cite{delfs2007} to obtain a new packet $P$. One possibility is to use an error correcting code such as \cite{moldovyan2017} in order to overcome errors due to noisy channels or imperfect photon detectors. For optimal security, a one-time pad symmetric key whose length is at least as long as the message should be used. Other keys may not guarantee unconditional security of the transmission. Let us assume that we use Vernam's One-Time Pad \cite{vernam1926}: $P=K \oplus C$, where $\oplus$ indicates the bitwise XOR operation.

\item Alice adds a small number of random bits at random locations of $P$ as a redundancy check to obtain a new packet $T$.
\item Inside the quantum encoder, Alice encodes each bit of $T$ by two qubits in the computational basis $\{|0\rangle,\ |1\rangle\}$ according to the logical bit before applying a Hadmard gate to one of the two qubits selected randomly (see Figure 2-a).
 
 \item  Bob receives each pair of qubits and either applies a Hadamard gate to the two qubits or not randomly and records his measurements for each pair, as in Figure 2-b.
 
 \item Bob sends to Alice over the classical channel the indices of the pairs where his measurements outcomes agree. These are the bits of $T$ which Bob could decode independently of Alice.
 \item Alice sends to Bob over the classical channel her choices of the basis for the other pairs. Bob uses this information to decode the rest of $T$.
\item Alice sends to Bob the indices of the redundant bits added to $P$ and they compare their values of these bits over the classical channel. If the number of discrepancies between them is higher than a certain threshold determined by the noise of the channel, they conclude that an eavesdropper is intercepting the transmission and the transmission is aborted. Otherwise, $P$ is recovered from $T$ by removing the redundancy bits.

\item Alice proceeds with the transmission of the key. The key is fed into the quantum encoder and steps 4-7 are repeated for $K$.   

\item Bob uses $K$ to decrypt $P$ in order to obtain $C$; $C=P\oplus K$. He computes the hash value from $M$ and compares it with the received one ($S$). In case of discrepancies, they conclude that either the channel is too noisy and the errors have corrupted the message/key or the whole transmission is compromised.

 \end{enumerate}

\section{DSQC protocol with entanglement}

Here, we introduce a second protocol which is similar to the one presented in the previous section in terms of the classical preprocessing, but differs in the quantum encoding stage.
 In this protocol, the message bits (the logical bits) are the control qubits of the the second Z-gate in the circuit used by Alice as shown in in Figure 2-c.  Alice also has the freedom to randomly send either two entangled qubits or two non-entangled qubits depending on the random control qubit of her two-qubit controlled-Z gate (the third Z-gate in the Alice's circuit). In the first case, when the controlled-Z gate is not enabled, she encodes the `1' by the  state   $\frac{1}{\sqrt{2}}(|-0\rangle-|+1\rangle)$ and `0' by the state $\frac{1}{\sqrt{2}}(|+0\rangle-|-1\rangle)$. The two states are verified to be entangled using the Peres-Horodecki criterion \cite{leinaas2006}. On the other hand, when Alice does not enable the controlled-Z gate, she encodes the `1' by the  state  $|--\rangle$ and `0' by the  state  $|+-\rangle$. We note that both states are non-entangled and the message in this case is encoded by the first qubit only, while the second qubit is redundant. The two-qubit controlled Z-gate can be implemented using the standard Toffoli and controlled-Z gates as shown in Figure 2-e.

Bob, on his side, also has the freedom to insert a controlled-Z gate before he applies a Hadamard gate on the first qubit as shown in in Figure 2-d. If both Alice and Bob enable their controlled-Z gate, the two-qubit state directly before the measurement of Bob will be 
$\frac{1}{\sqrt{2}}(|10\rangle-|01\rangle)$ for  `1' and $\frac{1}{\sqrt{2}}(|00\rangle-|11\rangle)$ for `0', i.e., Bob can distinguish the two message bits by detecting whether the two  measurements agree or not. Interestingly, if neither Alice nor Bob inserts their controlled-Z gate, the two-qubit state directly before the measurement of Bob will the same as in the previous case and Bob will be using the same rule. On the other hand, if the choices of Alice and Bob don't agree, the states of the two qubits just before the measurement of Bob will be  $|1-\rangle$ for `1' and $|0-\rangle$ for `0', i.e., Bob will be looking at the measurement of the first qubit only to decode the logical bit. Therefore, Alice should communicate her random choices of the two-qubit controlled Z-gate generated by the random number generator to Bob at the end of the transmission of each packet.

\section{Security analysis and discussion}

Let us imagine a typical eavesdropping scenario and analyze the quantum bit error rate (QBER) caused by it, assuming that perfect photon detectors are used by all sides. Let us consider first the quantum encoder without entanglement.  A typical strategy Evan can follow after intercepting the two qubits is to behave as Bob by measuring the two qubits in the same basis, re-encoding them, as Alice would do, and sending them forward to Bob. This is called intercept-resend-attack. Let also us assume that Evan can listen to the classical channel without interrupting it. That indicates he will know the correct values of the logical bits being sent by Alice after the public exchange between Alice and Bob corresponding to these cases as well as the cases in which he could decode the logical qubits on his own. The two cases combined amount to 75\% of all bits, i.e., the mutual information between Alice and Evan ($I_{AE}$) is 0.75. But what about the errors his intervention will cause at the side of Bob? Since Bob randomly applies the Hadamard gates to both qubits, he will get the same values measured by Evan at half the time and completely random values at the other half. In the last case, the random values will cause errors with probability 50\%. Therefore, the bit error rate, assuming a noiseless channel, caused by the intervention of Evan is 25\%, similar to the QBER of BB84 protocol for the same kind of attack. More complex attack scenarios may result in lower QBER. We show in Fig. 3-a and 3-b QBER and $I_{AE}$ obtained numerically by simulating Evan's attacks and using packets of length 10000 bits. The rate of Evan's intervention $\epsilon$ varies from 0 to 1. We can see that at $\epsilon=1$, i.e., Evan intercepts all the qubits, we obtain the theoretical predictions justified earlier, QBER=0.25 and $I_{AE}=0.75$.


\begin{table*}[ht]

\centering
\begin{tabular}{@{}llclllll@{}}
\toprule
                           & Elsayed I             & \multicolumn{1}{l}{Elsayed II} & Shimizu \cite{shimizu1999}               & Beige  \citep{beige2002}                 & Rong I \cite{rong2020-I}               & Rong II \cite{rong2020-II}               & Zou \cite{zou2014}                   \\ \midrule
Entangled photons          & \multicolumn{1}{c}{}  & \checkmark                              & \multicolumn{1}{c}{\checkmark} &                       &                       & \multicolumn{1}{c}{\checkmark} &                       \\
Single-way quantum channel & \multicolumn{1}{c}{\checkmark} & \checkmark                              & \multicolumn{1}{c}{\checkmark} & \multicolumn{1}{c}{\checkmark} &                       &                       &                       \\
Classical Bob              &                       & \multicolumn{1}{l}{}           &                       &                       & \multicolumn{1}{c}{\checkmark} & \multicolumn{1}{c}{\checkmark} & \multicolumn{1}{c}{\checkmark} \\
Quantum memory             &                       &  &                       &                       &                       &                       & \multicolumn{1}{c}{\checkmark} \\ \bottomrule
\end{tabular}
\caption{ \label{table-1}   Comparison between different deterministic quantum communication protocols, including the two protocols presented in this paper. The criteria are whether the protocol use single photons or entangled photons, single-way quantum channel or two-way quantum channel, whether Bob uses quantum operations on the received photons before the measurement in the computational basis or not and whether a quantum memory used to store the photons is needed or not. }
\end{table*}

Since single photon sources are often not ideal, an emitted light pulse supposed to contain one photon can carry more or less than a single photon. This leaves room for a smart eavesdropper to perform a photon-number-splitting attack, whereby he splits the light pulse if it carries more than one photon, and stores one photon at his disposal \cite{brassard2000}. Evan can then listen to the classical communication between Alice and Bob in order to get any clue about the right measurement basis to perform on the stolen photon. Let us analyze the effect of this attack on our DSQC scheme without entanglement. If Evan succeeds to get duplicates of every pair of qubits sent to Bob, he will wait for the classical communication between Alice and Bob to take place and listen to the choices Alice sends to Bob at certain locations. This scenario corresponds to 50\% of all transmitted qubit pairs. For the other 50\% where Bob could decode the logical bit on his own without the need to communicate with Alice, Evan will have to do exactly the same as Bob did; that is to measure the two qubits with a similar, but random basis. In half the cases, Evan will get similar measurement outcomes, thus decodes the logical bit correctly. Therefore in this eavesdropping scheme, Evan will be able to decipher overall 75\% of all the message. This is quite a huge ratio, but it also assumes ideal circumstances at the side of Evan, such as the ability to store photon pairs for a long time and the ability to get duplicates of every photon pair transmitted from Alice to Bob.

Let us now consider the same intercept-resend attack for the quantum encoder with entanglement. As before, Evan will intercept the two qubits, try to extract the maximum information, and resend them to Bob. While posing as Bob, Evan will randomly enable the controlled-Z gate,  record his measurements and then generate a quantum state of two qubits before he sends them to Bob. Evan will not, however, be able to decode the logical qubits before the end of the packet transmission when Alice communicate her choices for the two-qubit controlled-Z gate to Bob. He will then be able to decode all the bits with 100\% success probability, but at what cost? Since Evan has to resend the qubits to Bob on time, before he succeeds in decoding the logical bits, he will have to assume random values for the message bits and Alice's random number generator bits and use these random bits in the circuit that generates the states he sends to Bob, causing unavoidable errors. We show in Fig. 3-c the numerical simulation of QBER for this attack and we notice that for a 100\% intervention rate, QBER=0.5. 

Another attack strategy by Evan that will surprisingly leave no trace, i.e., cause no errors at the side of Bob, is to measure the first qubit only in the rectilinear basis, $\{|-\rangle,|+\rangle$. In this case, Evan will always decode the logical bits sent via non-entangled states correctly, while he will only decode the cases where entanglement is used with 50\% probability of success. Thererfore, the information Evan will retrieve in this case will be only 75\% of the packet. This information will not be helpful to Evan since the key will also be encoded by the quantum encoder, therefore, Evan will not gain full access to either the key or the encrypted message. We performed a numerical simulation for $I_{AE}$ of this attack while varying the rate of Evan's intervention $\epsilon$ from 0 to 1. The results are presented in Fig. 3-d.


There is another class of secure direct communication protocols where, unlike other protocols such as the ones presented in this paper, Bob is not a fully quantum agent. In these semi-quantum protocols such as \cite{zou2014,rong2020-I,rong2020-II}, Bob does not have to perform complex quantum operations other than measuring the received photons in the computational basis or reflecting the photons back to Alice. In Table 1, we give a comparison between the two protocols presented in this paper, the semi-quantum protocols \cite{zou2014,rong2020-I,rong2020-II}, and two of the earliest DSQC protocols, namely the protocol of Beige et. al. \cite{beige2002} where single photons are used and the protocol by Shimizu et. al. \cite{shimizu1999} which uses entangled photons. The later protocol also intersperses the message with random check bits for detecting eavesdropping like ours. We compare between these seven protocols based on whether the protocol uses single photons or entangled photons, single-way quantum channel or two-way quantum channel, whether Bob uses quantum operations on the received photons before the measurement in the computational basis or not and whether a quantum memory used to store the photons is needed or not.

In conclusion, it was shown that by combining the methods of classical cryptography and  quantum encryption we can find new protocols for deterministic secure quantum communication that encodes both the key and the message with the same quantum algorithm. In the proposed scheme, we send the key through the quantum channel with the same quantum encryption technique we use for the message, therefore it represents an intermediate case between QSDC where no key at all is used and conventional DSQC techniques where a key is sent over the classical channel. While for the quantum one-time pad protocol \cite{deng2004} a security check is performed before the message is sent, here, we perform the check concurrently while sending the encrypted message. The proposed schemes do not require a two-way quantum channel nor a quantum memory. They are in principle similar in nature to the deterministic algorithm proposed in \cite{beige2001,beige2002}, which use  the spatial and polarization degrees of freedom of single photons,  where the message is encrypted by Alice using a secret crypto key before being sent to Bob and also random control bits are inserted that aim to detect eavesdropping.  A full security proof and the analysis of more complex attack strategies and the effects of imperfect detectors and channel losses are required to ensure the security and practicality of the proposed protocol.


The author thanks Prof. Mark Hillery for the hospitality of Hunter College of CUNY where this work was initiated.

\bibliographystyle{andp2012}
\bibliography{qsdc}


\newpage
\begin{figure*}[] \setlength{\unitlength}{0.1cm}

\begin{picture}(125 , 130 ) 
{

\put(0, 0)  {\includegraphics[ scale=0.5]{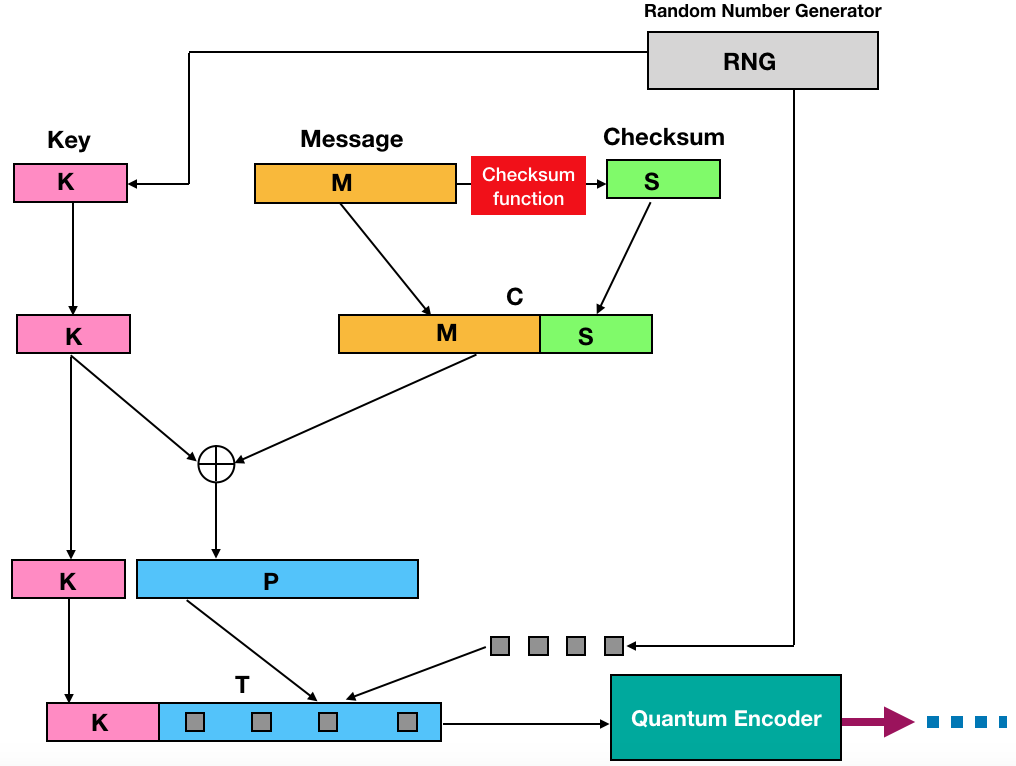}}

}
\end{picture} 
\caption{ \label{preprocessing} The preprocessing block diagram of a classical message before it is fed into a quantum encoder. A hash value (checksum) is added to the message before it is being encrypted with a random key. Random redundancy check bits are inserted into the the encrypted message at random locations. }
\end{figure*} 
 \normalsize

\newpage

 \begin{figure*}[] \setlength{\unitlength}{0.1cm}

\begin{picture}(165, 145 ) 
{
\put(50, 0)  {\includegraphics[ scale=0.5]{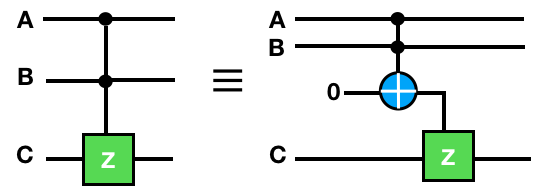}}
\put(-3, 39)  {\includegraphics[ scale=0.5]{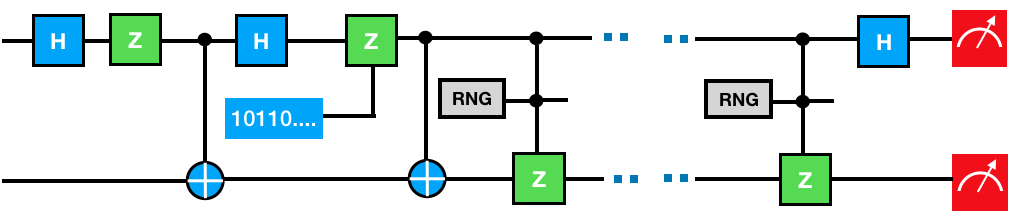}}
\put(93, 90.5)  {\includegraphics[ scale=0.5]{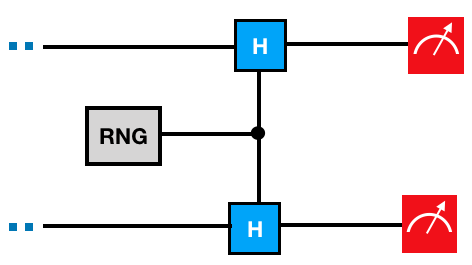}}
\put(-3, 90)  {\includegraphics[ scale=0.5]{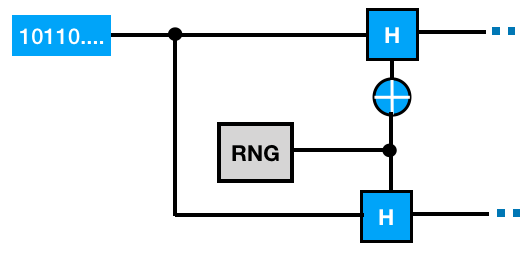}}
\put(75, 109)  {\includegraphics[ scale=0.4]{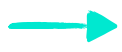}}
\put(99, 55)  {\includegraphics[ scale=0.4]{arrow.png}}

\put(-5,138) {\bf (a) }
\put(-5,79) {\bf (c) }
\put(133,79) {\bf (d) }
\put(133,138) {\bf (b) }
\put(88,33) {\bf (e) }
\put(-4,63) {$|0\rangle$ }
\put(-4,38) {$|0\rangle$ }
}
\end{picture} 

\caption{ \label{encoder} (a, b)  The circuit diagram of the quantum encoder and decoder for the DSQC protocol that does not use entanglement. Every bit is encoded by two qubits. One random qubit is encoded in the computational basis $\{ |0 \rangle,\  |1   \rangle \} $ and the other one in the Hadamard basis $\{  \frac{1}{\sqrt{2}}(|0 \rangle \pm  |1   \rangle) \} $. The receiver, on the other hand, measures the two qubits in either the computational basis or the Hadamard basis in a random manner. The random choices are determined by a random number generator (RNG). (c, d)  The circuit diagram of the quantum encoder and decoder for the DSQC protocol that uses entangled qubits. `1' and `0' are encoded by two qubits in the states $\frac{1}{\sqrt{2}}(|-0\rangle-|+1\rangle)$ and $\frac{1}{\sqrt{2}}(|+0\rangle-|-1\rangle)$  respectively or the  states  $|--\rangle$ and $|+-\rangle$ respectively depending on the values of the random qubits generated by the random number generator (RNG) at the side of Alice. (e) An implmentation of the two-qubit controlled-Z gate used by Alice and Bob in (c,d) using a Toffoli gate and a controlled-Z gate.  }
\end{figure*}

\newpage

 \begin{figure*}[] \setlength{\unitlength}{0.1cm}

\begin{picture}(165, 90 ) 
{
\put(30, 45)  {\includegraphics[ scale=0.45]{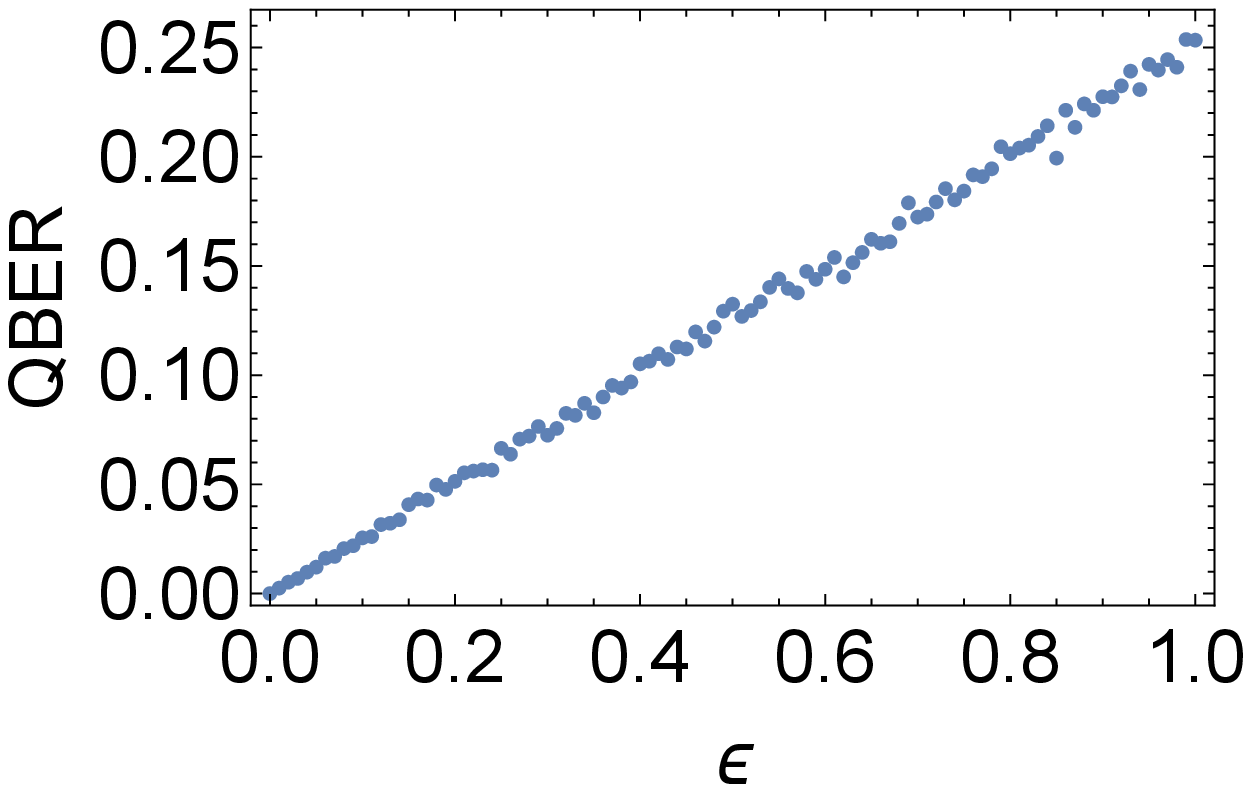}}
\put(90, 45)  {\includegraphics[ scale=0.45]{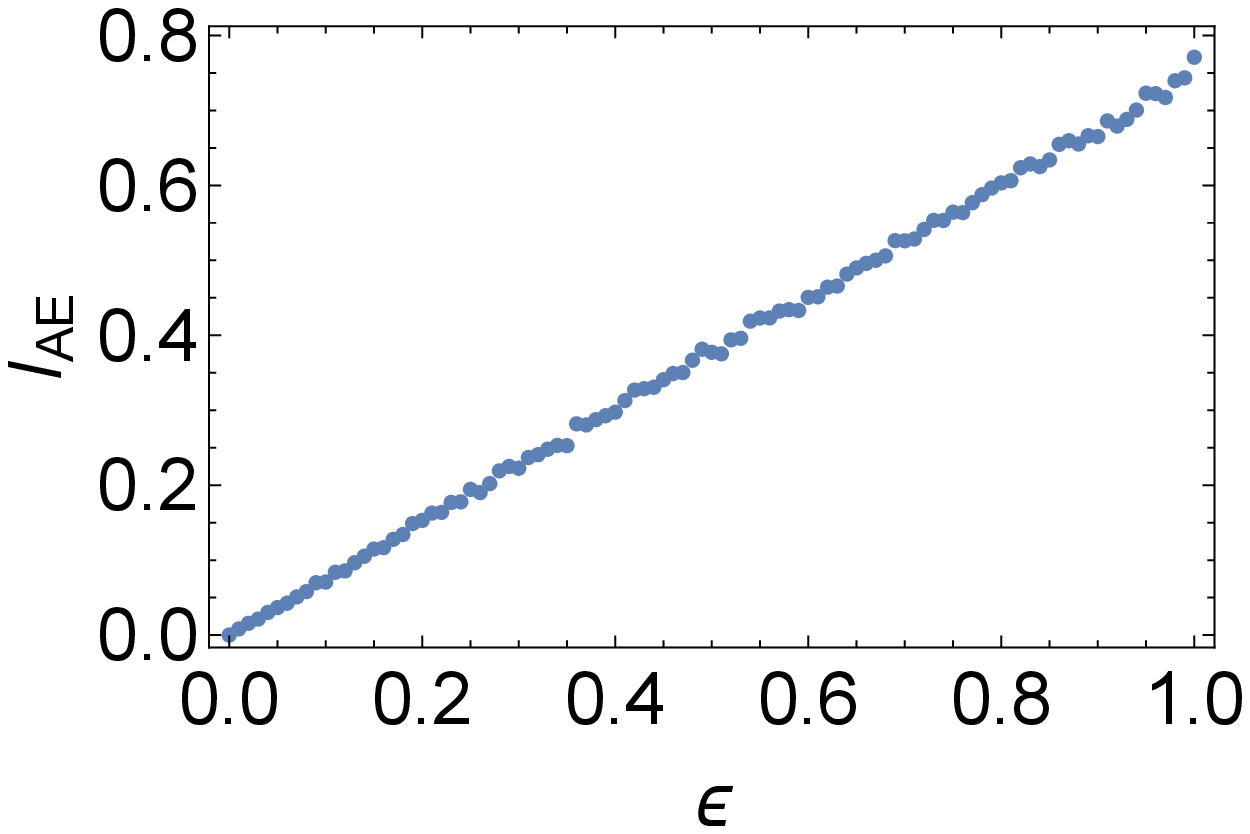}}
\put(30,0)  {\includegraphics[ scale=0.45]{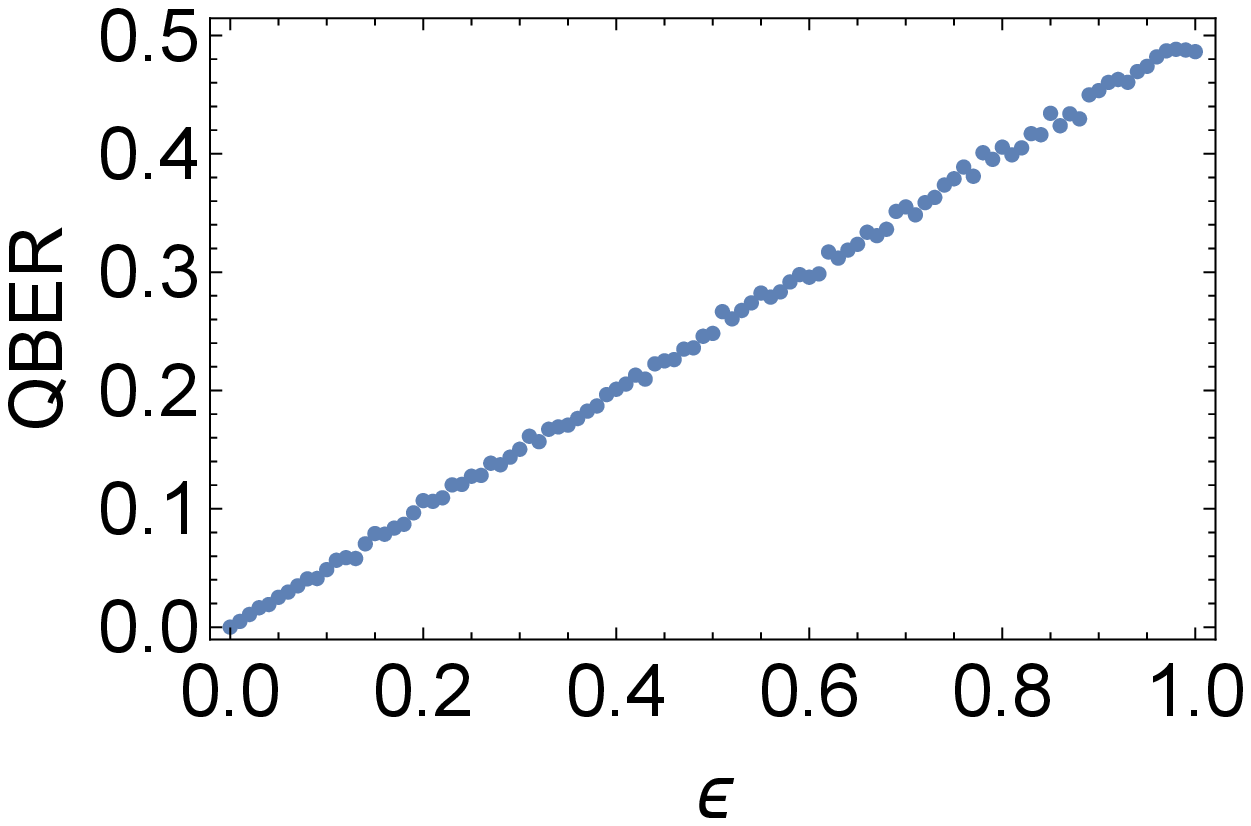}}
\put(90,0)  {\includegraphics[ scale=0.45]{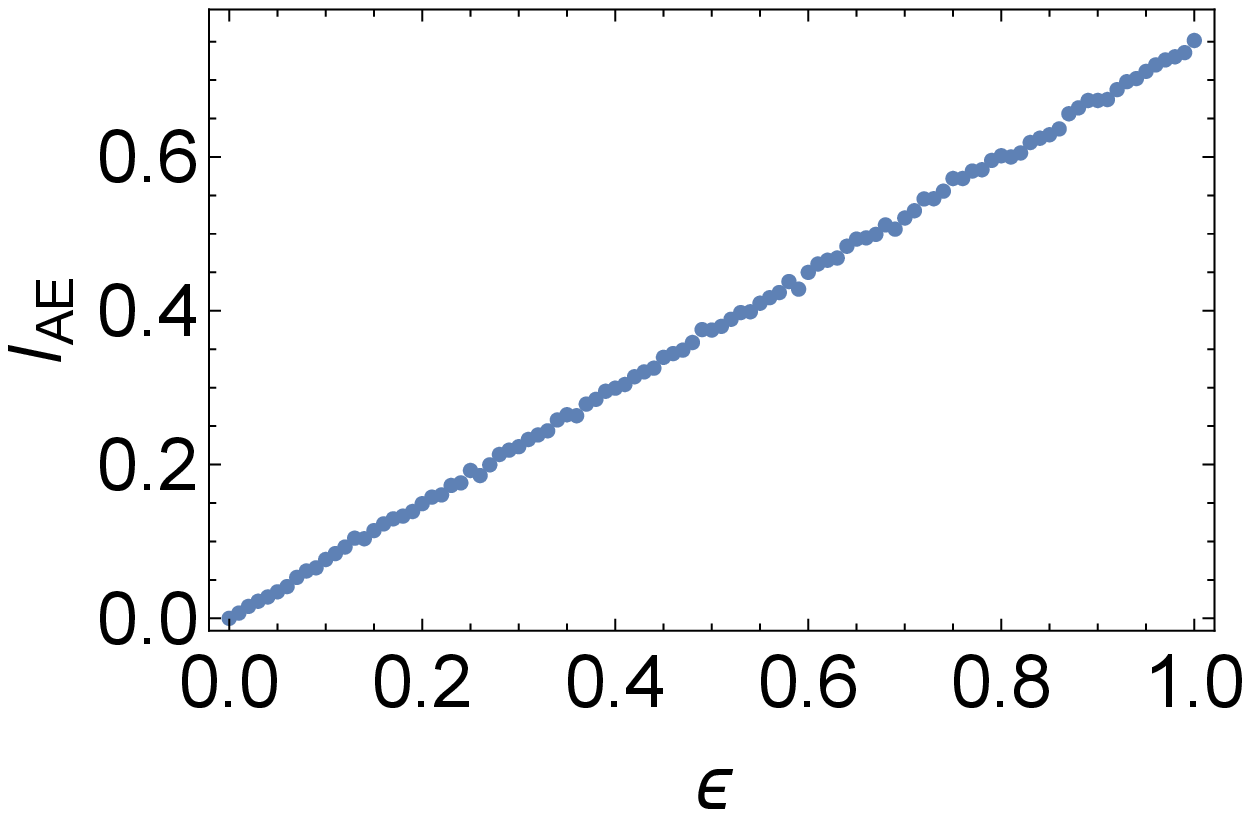}}

\put(32,85) {\bf (a) }
\put(90,85) {\bf (b) }
\put(30,40) {\bf (c) }
\put(90,40) {\bf (d) }
}
\end{picture} 

\caption{ \label{simu} Numerical simulation of the proposed quantum encoders for various intercept-resend attack strategies by Evan and for different values of Evan's intervention rates. (a) and (b) the quantum bit error rate (QBER) and the mutual information between Alice and Evan ($I_{AE}$) for the same intercept-resend attacks on the quantum encoder without entanglement.  (c) and (d) QBER and $I_{AE}$ for two different attack strategies on the quantum encoder with entanglement. In the first attack (c) Evan can decode 100\% of the logical bits while in the second attack (d) Evan can cause no no errors in the logical bits received by Bob.       }
\end{figure*}


 \normalsize

\end{document}